\documentclass[prl,aps,showpacs,twocolumn,superscriptaddress]{revtex4}
\usepackage{graphicx,bm}
\usepackage{amssymb}
\usepackage{epsfig}
\usepackage{epsf}
\usepackage{amsmath}
\usepackage[usenames]{color}

\begin{document}

\title{Spin-wave amplification and lasing driven by inhomogeneous spin transfer torques}

\author{R.J. Doornenbal}\affiliation{Institute for Theoretical Physics, Utrecht
	University, Leuvenlaan 4, 3584 CE Utrecht, The Netherlands}


\author{A. Rold\'an-Molina}
\affiliation{Universidad de Aysén, Calle Obispo Vielmo 62, Coyhaique, Chile}

\author{A.S. Nunez}
\affiliation{Departamento de F\'isica, Facultad de Ciencias F\'isicas y Matem\'aticas, Universidad de Chile, Casilla 487-3, Santiago, Chile}

\author{R. A. Duine}

\affiliation{Institute for Theoretical Physics, Utrecht
University, Leuvenlaan 4, 3584 CE Utrecht, The Netherlands}

\affiliation{Department of Applied Physics, Eindhoven University of Technology, P.O. Box 513, 5600 MB Eindhoven, The Netherlands}

\date{\today}

\begin{abstract}
We show that an inhomogeneity in the spin-transfer torques in a metallic ferromagnet under suitable conditions strongly amplifies incoming spin waves. Moreover, at nonzero temperatures the incoming thermally occupied spin waves will be amplified such that the region with inhomogeneous spin transfer torques emits spin waves spontaneously, thus constituting a spin-wave laser. We determine the spin-wave scattering amplitudes for a simplified model and set-up, and show under which conditions the amplification and lasing occurs. Our results are interpreted in terms of a so-called black-hole laser, and could facilitate the field of magnonics, that aims to utilize spin waves in logic and data-processing devices. 
\end{abstract}


\maketitle

\def\bx{{\bm x}}
\def\bk{{\bm k}}
\def\bK{{\bm K}}
\def\bq{{\bm q}}
\def\br{{\bm r}}
\def\bp{{\bm p}}
\def\bM{{\bm M}}
\def\bs{{\bm s}}
\def\bB{{\bm B}}
\def\bj{{\bm j}}
\def\bF{{\bm F}}
\def\id{{\rm d}}

\def\br{{\bm r}}
\def\bv{{\bm v}}

\def\half{\frac{1}{2}}
\def\args{(\bm, t)}

\renewcommand{\r}{\ensuremath{\rho}}
\renewcommand{\a}{\ensuremath{\alpha}}
\renewcommand{\b}{\ensuremath{\beta}}
\renewcommand{\c}{\ensuremath{\gamma}}
\renewcommand{\d}{\ensuremath{\delta}}
\newcommand{\D}{\ensuremath{\Delta}}
\renewcommand{\o}{\ensuremath{\omega}}
\renewcommand{\O}{\ensuremath{\Omega}}
\renewcommand{\l}{\ensuremath{\lambda}}
\newcommand{\s}{\ensuremath{\sigma}}
\newcommand{\p}{\ensuremath{\psi}}
\renewcommand{\t}{\ensuremath{\theta}}
\newcommand{\ta}{\ensuremath{\tau}}
\newcommand{\e}{\ensuremath{\epsilon}}
\renewcommand{\k}{\ensuremath{\kappa}}

\newcommand{\lb}{\ensuremath{\left(}}
\newcommand{\rb}{\ensuremath{\right)}}
\newcommand{\lbb}{\ensuremath{\left[}}
\newcommand{\rbb}{\ensuremath{\right]}}
\newcommand{\la}{\ensuremath{\left\langle}}
\newcommand{\ra}{\ensuremath{\right\rangle}}

\newcommand{\al}[1]{\begin{align}#1\end{align}}
\newcommand{\all}[1]{\begin{align*}#1\end{align*}}
\newcommand{\intR}{\ensuremath{\int_{-\infty}^\infty}}
\newcommand{\dd}{\partial}
\newcommand{\na}{\nabla}
\newcommand{\hb}{\hbar}
\newcommand*\conj[1]{\overline{#1}}
\renewcommand{\Re}{\mathrm{Re\,}}
\renewcommand{\Im}{\mathrm{Im\,}}

\renewcommand{\SS}{\bm S}
\newcommand{\vv}{\bm v}
\newcommand{\kk}{\bm k}
\newcommand{\ee}{\mathcal{E}}
\newcommand{\be}{\bm{ \mathcal{E}}}
\newcommand{\ts}{\t_{SH}}
\newcommand{\n}{\bm{n}}
\renewcommand{\v}{\bm{v}_s}
\newcommand{\pc}{\psi^{*}}
\newcommand{\oM}{\o_{\text{max}}}
\newcommand{\om}{\o_{\text{min}}}
\newcommand{\xiM}{\xi_{\text{max}}}

{\it Introduction.} --- Spin waves are collective excitations in magnetically-ordered materials. At the semi-classical level, spin waves in ferromagnets correspond to a wave-like pattern of precessing spins in which the relative phase of the precession of two spatially separated spins is determined by the ratio of their distance to the wavelength of the spin wave. When the exchange interactions dominate, the spin precession is circular. Anisotropies and dipolar interactions, however, generically lead to elliptically-precessing --- in short, elliptical --- spin waves.

Though spin waves are neutral excitations, they are able to transfer angular momentum. Magnonics \cite{kruglyak2010, chumak2017} is named after the quasi-particle, the magnon, associated with a spin wave. This field has the ultimate goal of controlling and manipulating spin waves to the point that they can be used to realize energy-efficient data-processing and logic devices. One hurdle to realize technology based on spin waves is that they have a finite lifetime as a result of processes that lead to loss of spin angular momentum and relax the magnetization. Hence, experimental progress has been nearly exclusively made using a unique low magnetic-damping material: the complex magnetic insulator Yttrium Iron Garnet (YIG), thereby limiting the process as it is difficult to fabricate and pattern at high quality in reduced dimensions. 

The relaxation of spin waves can be counteracted by injection of spin angular momentum to compensate for the losses. This has been demonstrated in YIG/Pt-based material systems \cite{demidov2014}, in which a charge current, driven through the Pt and tangential to the interface with YIG, excites a --- via the spin Hall effect \cite{sinova2015} --- spin current that is absorbed by the magnetization in the magnetic insulator. A similar result has been obtained with the magnetic metal permalloy and Pt \cite{gladii2016}. In a different implementation \cite{padron2011}, the spin current was injected by a thermal gradient via the spin Seebeck effect \cite{uchida2010}, and increased spin-wave propagation lengths were also observed. In these examples, the amplitude enhancement of the spin waves is proportional to the applied charge current or temperature gradient, which may be a limiting factor in case the damping that needs to be overcome is large, or because of the associated heating. 

\begin{figure}
	\includegraphics[width=9cm]{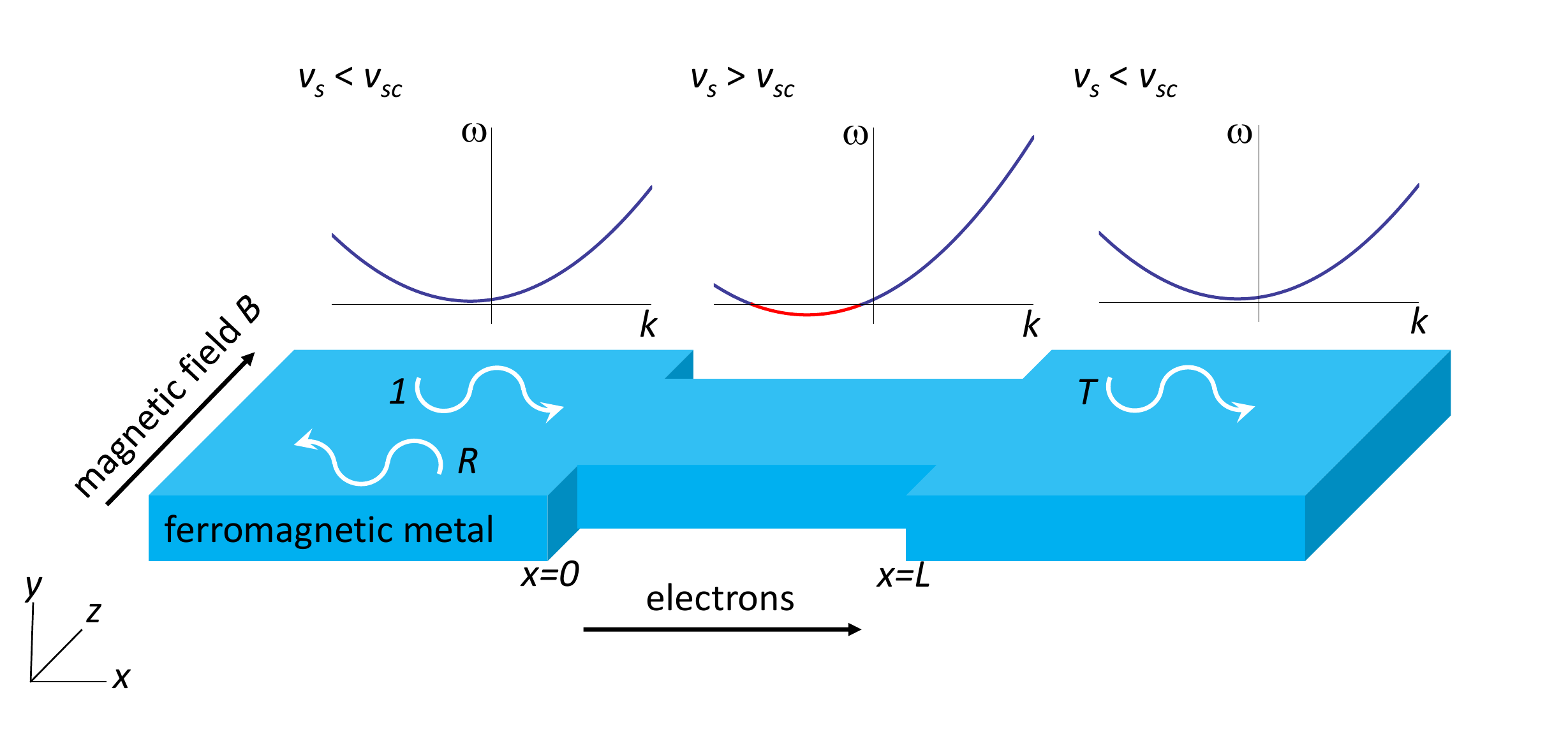} 
	\caption{The set-up that is considered in this Letter: A ferromagnetic wire with magnetization saturated in the $z$-direction is subjected to a current density driven in the long direction of the wire. The wire has an indent such that the current density in this narrow region is larger than in the wider parts of the wire. For large enough currents, the resulting current-induced spin-wave Doppler shift pushes the spin-wave energies indicated in red to negative values in the narrow part of the wire while the spin-wave energies in the wide parts of the wire remain positive. The spin-wave frequency $\omega$ is sketched as a function of the wave number $k$ in the three different regions.}
	\label{fig:setup}
\end{figure}

In this Letter, we propose a different way to amplify spin waves. It makes use of the spin transfer torques that arise in the bulk of ferromagnetic metals as a result of the interaction between the spin-polarized electronic current and the magnetization \cite{ralph2008}. The basic set-up we consider is sketched in Fig.~\ref{fig:setup}. It consists of a ferromagnetic metallic wire with a constriction. A charge current driven through the wire will have a larger current density in the narrow part of the wire, as compared to the wider parts. As a result, the velocity that characterizes the spin transfer torques and determines, for example, the current-induced spin-wave Doppler shift \cite{Vlaminck2008} will be larger in the narrower part of the wire. For sufficiently large current densities, the Doppler shift will make the spin-wave energies and frequencies in the narrow part of wire negative (indicated in red in Fig.~\ref{fig:setup}), while in the wider parts of the wire they remain positive. Moreover, in the case of a finite spin-wave ellipticity, the spin waves with positive and negative energy are coupled in the regions where the width of the wire changes. As a result of this coupling, spin waves can be created simultaneously in the wide and narrow parts of the wire without changing the total magnetic energy. For the spin-wave modes that bounce back and forth in the narrow part of the wire, the coupling between negative-energy and positive-energy spin waves can be resonant. We show that this leads to an enormous enhancement of the transmission and reflection amplitudes for spin waves that are scattered off the narrow part of the wire and that fulfil the resonance condition. Moreover, this enhancement is so large that the thermally-excited spin waves that fulfil the resonance condition will be amplified and emitted from the narrow part of the wire and overwhelm all other spin waves, thus constituting a spin-wave laser. Below, we discuss our set-up in more detail and provide analytical results to underpin our proposal. 

{\it Model and set-up.} --- We consider a ferromagnetic metal subject to an external field $B$ in the $z$-direction, and charge current ${\bm j}$. The magnetization direction $\n (\bx,t)$ obeys the Landau-Lifshitz-Gilbert equation with spin transfer torques that is given by \cite{ralph2008}
\begin{equation}
\label{eq:llgwithstts}
	(\dd_t + \v \cdot \na )\n = \n \times \bm H_{\mathrm{eff}} - \a \n \times \lb \dd_t + \frac{\b}{\a} \v \cdot \na \rb \n~.
\end{equation}
This equation describes precession of the magnetization around the effective field $\bm H_{\mathrm{eff}} = -\delta E/\hbar \delta \n$, where $E$ is the magnetic energy functional. We consider a generic energy functional that incorporates exchange, anisotropy, and the external field and is given by
\begin{equation}
E = \int \frac{dV}{a^3} \lbb -\frac 1 2 J \n \cdot \na^2 \n + \frac 1 2 K_x n_x^2 + \frac 1 2 K_y n_y^2 - B n_z\rbb~.
\end{equation}
In this expression, $a$ is the size of the unit cell, $J$ is the exchange constant, and $K_x$ and $K_y$ are anisotropy constants. In case that the anisotropy constants are equal, the spin waves are circular. Due to shape and crystalline anisotropy the spin wave are typically elliptical, corresponding to the case that $K_x \neq K_y$. 

The spin transfer torques in the Landau-Lifshitz-Gilbert equation~(\ref{eq:llgwithstts}) are characterized by the velocity $\v =  -g P \mu_B {\bm j}/2eM_s$, that is proportional to the current density and further determined by the current polarization $P$, the Land\'e factor $g$, the Bohr magneton $\mu_B$, the elementary charge $e$, and the saturation magnetization $M_s$. The adiabatic spin transfer torque appears on the left-hand side of Eq.~(\ref{eq:llgwithstts}), whereas the non-adiabatic spin transfer torques appears on the right-hand side and is parametrized by the dimensionless constant $\beta \ll 1$. The Gilbert damping constant $\alpha \ll 1$ determines the rate of decay of the magnetization direction. 

We take $K_x,K_y,B>0$ so that the equilibrium direction of magnetization is the $z$-direction.  Linearizing around this equilibrium direction yields the dispersion relation 
\begin{equation}
\label{eq:dispfull}
 \omega_{\bm k} - {\bm v}_s \cdot {\bm k } = \omega^0_{\bm k}-i\alpha \omega^0_\bk-  i (\alpha - \beta)  {\bm v}_s \cdot {\bm k}~,
\end{equation}
with 
\begin{equation}
\hbar \omega^0_\bk= \sqrt{ (\hb \o_0)^2 +2J\hb \o_0 k^2 + J^2 k^4- \D^2 }~,
\end{equation} 
the real part of the spin-wave dispersion in the absence of current, and where $\hb \o_0 \equiv B + (K_x+K_y)/2$ and $\D \equiv \left(K_y - K_x\right)/2$ and we assumed $\D > 0$ without loss of generality. This parameter is too some extent tunable by the wire geometry. In deriving the above dispersion relation we took $\v$ constant, but we will drop this assumption shortly. 

For $|\v| > v_{s,c} = \sqrt{2J/\hb^2} \sqrt{\hb \o_0 + \sqrt{(\hb\o_0)^2-\D^2}}$, the real part of the dispersion in Eq.~(\ref{eq:dispfull}) becomes negative. This signals an energetic instability as the system may lower its energy by creating negative-energy excitations. From now on, we assume that $\beta \approx \alpha$. This implies that the system remains dynamically stable even when it is energetically unstable, because small-amplitude fluctuations are damped out. This results from the imaginary part of the dispersion relation in Eq.~(\ref{eq:dispfull}) which remains negative when $\beta \approx \alpha$. 

In the remainder of this Letter we consider the system sketched in Fig.~\ref{fig:setup}, in which a local increase in the velocity $\v$ is accomplished by a narrow region in a wire of the metallic ferromagnet. Moreover, we assume that the current density is such that $\v$ is above the critical value $v_{s,c}$ in the narrow part of the wire, whereas it is below the critical value in the wider parts of the wire. The resulting local spin-wave dispersions are also sketched in Fig.~\ref{fig:setup} and correspond approximately to shifted parabolas. The negative-energy modes in the narrow part of the wire are indicated by the red dispersion curve.

{\it Scattering solutions.} --- We now proceed to construct spin-wave scattering solutions to the Landau-Lifshitz-Gilbert equation. We neglect in first instance the magnetization relaxation and put $\alpha=\beta=0$. We assume, moreover, that the transverse dimensions of the wire in the $y$ and $z$-direction are very small so that we may drop the dependence of $\n$ on $y$ and $z$, and may, moreover, take $\v = v_s (x) \hat x$. We further assume that the regions where the wire becomes wider and narrower are very small so that we may put $v_s \equiv v_L < v_{s,c}$ independent of $x$ in the wider part of the wire and $v_s \equiv v_M >v_{s,c}$ independent of $x$ in the narrow part of the wire. We take the narrow part of the wire between $x=0$ and $x=L$. As we discuss later on when we give an interpretation of our results, the assumptions on the transverse dimensions of the wire and on the position dependence of $v_s (x)$ are not essential but are made to facilitate the computations. 

To construct the scattering solutions it is convenient to introduce $\psi \equiv n_x - i n_y$. The linearized solutions of the Landau-Lifshitz-Gilbert equation are then given by 
\begin{equation}
\label{eq:wavefct}
	\p(x, t) = u(x) e^{-i \o t} - v^{*}(x) e^{i \o t}~,
\end{equation}
where $u(x)$ and $v(x)$ obey
\begin{widetext}
\begin{equation}
\label{eq:matrixeq}
	\begin{pmatrix}
		\hb \o + i \hb v_s (x) \na + J \na^2 - \hb \o_0 & -\D \\
		-\D & - \hb\o - i \hb v_s (x) \na + J \na^2 - \hb \o_0 \\
	\end{pmatrix} \begin{pmatrix}
	u (x) \\
	v (x) \\
\end{pmatrix}=0~. 
\end{equation}
\end{widetext}
Because we have taken $v_s (x)$ to be piecewise constant, the above equations for $u$ and $v$ are solved by a plane-wave {\it ansatz} in both the central narrow region of the wire, and in its wider parts. Scattering solutions are now constructed by taking a spin wave incoming from the left, and calculating its reflection and transmission amplitude, by matching at the position where $v_s$ changes.

The plane-wave solutions take the form
\begin{equation}
\label{eq:scattsols}
\begin{pmatrix}
	u (x) \\
	v (x) \\
\end{pmatrix} =  
\begin{pmatrix}
	F \\
	G \\
\end{pmatrix} e^{i k x}~,
\end{equation} 
where $F, G$ are complex coefficients.

\begin{figure}
	\includegraphics[width=8.5cm]{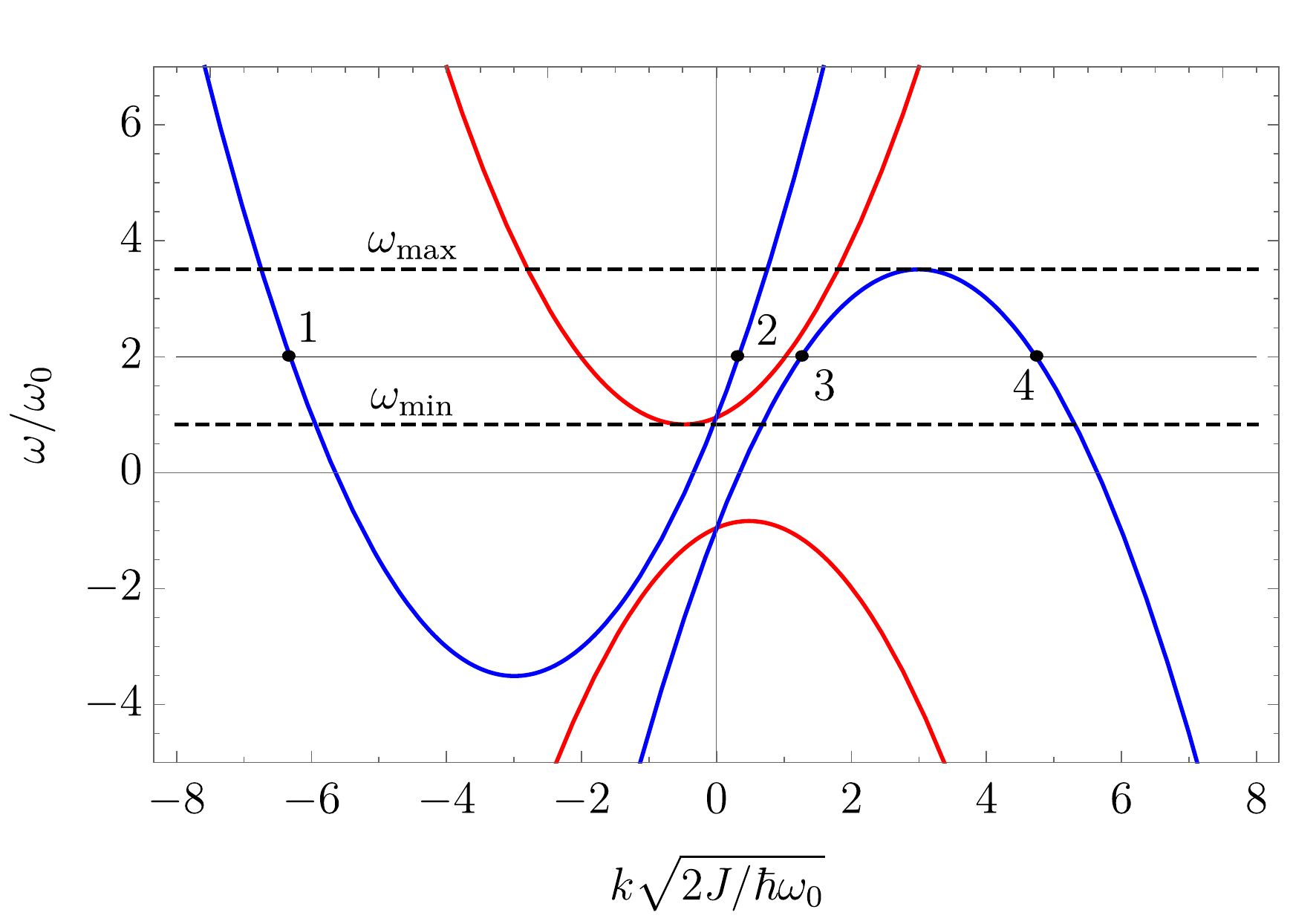} 
	\caption{Dispersion relation for $v_s < v_{s,c}$ (red) and $v_s > v_{s,c}$ (blue). Black dots mark the four wavenumbers corresponding to a generic value $\o < \oM$ in the $v_s > v_{s,c}$ sector. Parameter values: $v_L = 0.5 \sqrt{2J\o_0/\hb}, v_M = 3.0 \sqrt{2J\o_0/\hb}, \D = 0.3\, \hb \o_0 $. }
	\label{fig:dispersion}
\end{figure}

Using Eq.~(\ref{eq:matrixeq}) we obtain the dispersion relation
\al{
(\hb\o - \hb v_s k)^2 = (\hb \o_0)^2-\D^2+2J\hb \o_0k+J^2k^4~.
}
From now on we take $\o > 0$ without loss of generality.
At a given frequency $\o$, there are in general four (complex) values of $k$ that satisfy the dispersion. These are denoted by $k_i$, with $i=1, 2, 3, 4$ and are labelled according to  Fig.~\ref{fig:dispersion}.  Different $\o$ regimes need to be distinguished. Firstly, for $v_s < v_{s,c}$, the dispersion exhibits a gap $\o_{\text{min}}$. We require $\o \geq \o_{\text{min}}$ in order for scattering solutions to exist. For the regions where $v_s < v_{s,c}$ there are then two propagating modes with real $k$ and two growing/decaying modes with imaginary $k$. Secondly, for $v_s > v_{s,c}$, there exists a range of $\o$ from $\omega_{\rm min}$ to $\omega_{\rm max}$ within which there exist four real wave vectors $k$ that satisfy the dispersion relation. For $\o$ exceeding $\o_{\text{max}}$, two of the solutions for $k$ are real and two are imaginary. 

In what follows we will look at scattering solutions and hence assume $\o > \o_{\text{min}}$.  The coefficients of the respective growing modes for $x < 0$ and $x > L$ must vanish. In addition, we impose matching conditions at both jumps in $v_s$. The functions $u(x)$ and $v(x)$, as well as their first derivatives, are required to be continuous. This leads to a system of linear equations that can be solved for the reflected and transmitted amplitudes. Here, the reflection amplitude $R$ is defined as the ratio between the $F$-amplitudes [Eq.~(\ref{eq:scattsols})] of the incoming and reflected wave, whereas the transmission amplitude $T$ is defined as the same ratio but for incoming and transmitted wave. (One could also consider similar ratios of the $G$-amplitudes. This choice does not affect the location of the resonances.)

\begin{figure}
	\includegraphics[width=8.5cm]{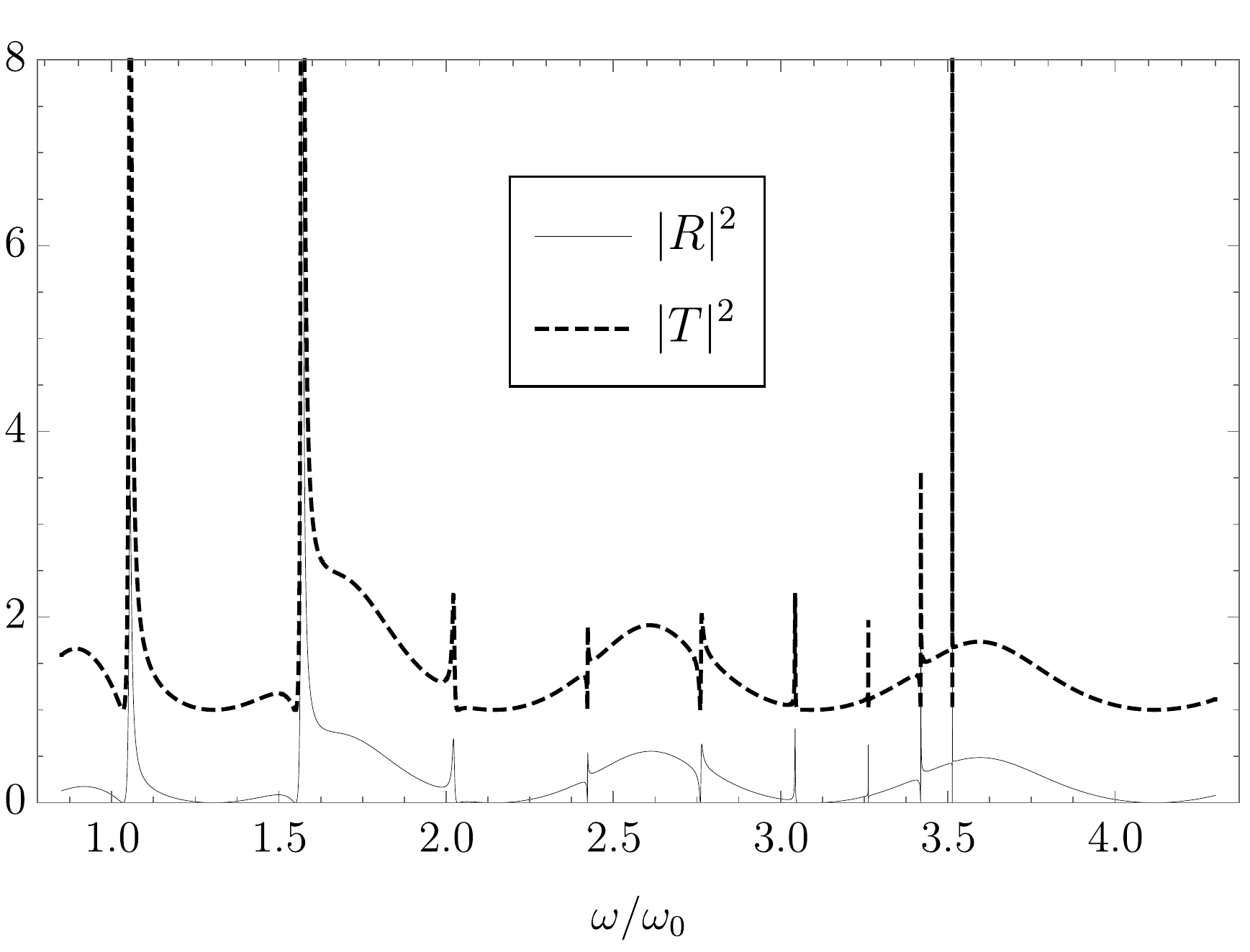} 
	\caption{Transmission (broken line) and reflection (solid line) probabilities off the spin-transfer torque inhomogeneity, see Fig.~\ref{fig:setup}, as a function of spin-wave frequency. Parameter values: $v_L = 0.5 \sqrt{2J\o_0/\hb}, v_M = 3.0 \sqrt{2J\o_0/\hb}, \D = 0.7\,\hb \o_0$.}
	\label{fig:transmission}
\end{figure}

\begin{figure}
	\includegraphics[width=8.5cm]{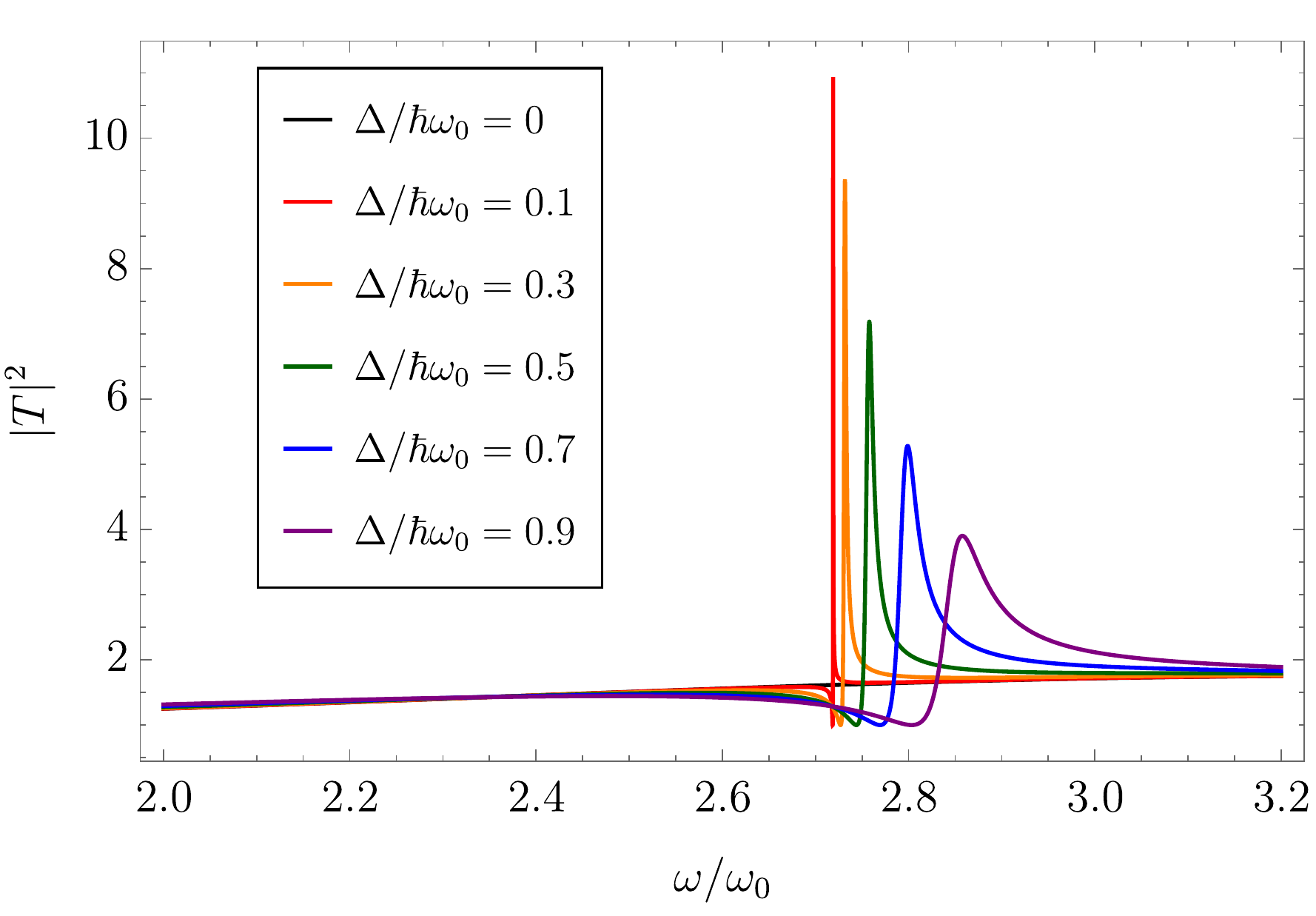} 
	\caption{Transmission probabilities near a resonance, for different $\D$. The resonance peak is sharper and higher for small $\D>0$, but is not present if $\D = 0$. Parameter values: $v_L = 0.5 \sqrt{2J\o_0/\hb}, v_M = 3.0 \sqrt{2J\o_0/\hb} $. }
	\label{fig:deltas}
\end{figure}

{\it Results.} --- In Fig.~\ref{fig:transmission} we show the results for the spin-wave transmission and reflection probabilities as a function of frequency, with the choice $L = 12 \sqrt{2J/\hb \o_0}$. For distinct frequencies both reflection and transmission are strongly enhanced. We refer to the peaks in the reflection and tranmission amplitudes as resonances. We have found that the resonance condition is well approximated by the equation
\al{ \label{eq:resonancecond}
|k_4 - k_3| = \frac{2\pi n}{L} + \mathcal{O}\lb \frac{1}{L^2} \rb
}
where $n = 1, 2, \ldots$. The first term of this equation has the physical interpretation that the counterpropagating waves corresponding to $k_3$ and $k_4$ interfere constructively between $x = 0$ and $x = L$. Near $\o = \oM$, the dispersion relation is approximately parabolic. This allows us to describe the resonant frequencies $\o_n$ with the simple approximate formula
\al{
\o_n = \oM - n^2 \Gamma^2,
}
where $n = 1, ..., \lfloor \sqrt{\oM-\om}/\Gamma \rfloor$ and $\Gamma \approx \pi \sqrt{J/\hbar L^2}  + \mathcal O\lb 1/L^2 \rb $. 
The value of $\Gamma$ is only weakly sensitive to the parameters $v_L, v_M, \D$. Varying the parameters $v_L, v_M$ affects the location of the resonances significantly only by shifting them all, via a change in the value of $\oM$. For $\o > \oM$, there are no resonances at all. This is explained by the fact that the wave vectors $k_3, k_4$ are complex in this case. In Fig.~\ref{fig:deltas} we have zoomed in on a single resonance peak for different values of $\D$. We find that smaller $\D$ results in higher and narrower peaks, but no resonances are present if $\D = 0$.

Given the strong enhancement of the reflection and transmission amplitudes, the occupation of the incoming spin waves by thermal fluctuations leads to emission of amplified spin waves thus forming a spin-wave laser. We denote with $u_n$ and $v_n$ the scattering solutions corresponding to the resonant energies for a spin wave incoming from the left. Furthermore, $u'_n$ and $v'_n$ denote the resonant scattering solutions for spin waves incoming from the right. These occur at the same energies as the scattering solutions for waves incoming from the left, which follows from the condition in Eq.~(\ref{eq:resonancecond}). We have that the magnetization direction, parametrized by $\psi (x,t)$, is approximately given by
\begin{eqnarray}
  \psi (x,t) \propto \sum_n n_B \left( \hbar \omega_n\right) \left[ u_n(x) e^{-i \o_n t} - v_n^{*}(x) e^{i \o_n t} \right. \nonumber \\
  \left.  u'_n(x) e^{-i \o_n t} - (v'_n)^{*}(x) e^{i \o_n t} \right]~.
\end{eqnarray}
If $1 < \sqrt{\oM-\om}/\Gamma < 2$, there is only a single resonance peak. Based on this, one can experimentally engineer spin-wave laser that emits spin waves at a single frequency. 

{\it Discussion and outlook.} ---
Our results can be interpreted as follows. The left transition region, i.e., where $v_s$ changes from $v_s < v_{s,c}$ to $v_s> v_{s,c}$, is a black-hole event horizon for spin waves \cite{jannes2011,roldan2017} coming in from the left (referring to Fig.~\ref{fig:setup}). The right transition from $v_s > v_{s,c}$ to $v_s < v_{s,c}$ is a white-hole horizon for spin waves coming from the right. Because the spin waves do not disperse linearly, these horizons are referred to as dispersive horizons \cite{chaline2013}. In the field of analogue gravity, such a pair of black-hole-like and white-hole-like event horizons, discussed in Ref.~\cite{roldan2017} for spin waves, is known to be able to give rise to so-called black-hole lasers \cite{corley1999,Steinhauer}, which exhibit the resonant amplification and lasing we have discussed in our specific set-up and model. Within this interpretation, the resonant amplification occurs as a result of the constructive interference of particle-hole coupling processes that arises at each horizon. (This particle-hole coupling gives rise to Hawking radiation in the quantum regime \cite{Hawking1974}.) For the system in Fig.~\ref{fig:setup}, the negative-energy modes are the holes, while the particles correspond to positive-energy modes.

The interpretation as a black-hole laser points to some essential ingredients for the spin-wave amplification and lasing. First of all, the negative-frequency and positive-frequency modes need to be coupled. This coupling occurs only for elliptical spin waves because these are a superposition of positive and negative frequencies, see Eq.~(\ref{eq:wavefct}). Secondly, though we have assumed a step-like current-density, our results are more general as any transition where $v_s$ goes from below (or above) to above (or below) $v_{s,c}$ will couple negative-energy and positive-energy modes and thus lead to amplification and lasing.  A unique ingredient of magnetic systems is the way the horizons are implemented, i.e., using electric current rather than flow of the spin waves themselves. This gives rise to the non-adiabatic spin-transfer torque, determined by the parameter $\beta$, which has no counterpart in other analogue gravity systems. The existence of these non-adiabatic spin transfer torque is crucial to make the system dynamically stable. 

Typical experimentally accessible values are $J \sim 10^{-39}$ Jm$^2$ and $B/\mu_B \sim K/\mu_B \sim 0.1-1$ T \cite{roldan2017}, so that $\omega_0 \sim 10-100$ GHz, and the length scale $\sqrt{J/\hbar \omega_0} \sim 10-100$ nm. This means that resonances should be visible for systems in the range $L \sim 10-1000$ nm, which is easily accessible experimentally. While we have in most of our treatment ignored magnetization relaxation (except for requiring that $\beta \approx \alpha$, which is a typical situation), our results remain valid provided the spin-wave coherence length is much long than $L$. This translates to the condition that $1/(\alpha k_i) \gg L$. This condition is easily satisfied given that $\alpha \ll 1$. For the above-mentioned typical values of anistropies and external fields, the current density corresponding to $v_{s,c}$ is on the order of $10^{12}$ A$/$m$^2$ \cite{roldan2017}. Though large, this current density is reached often, e.g., in experiments using pulsed current-driven domain-wall motion \cite{ralph2008}. This critical current density may be lowered in systems that involve Dzyaloshinskii-Moriya interactions. In such systems, the spin-wave dispersion has a linear part in the dispersion without current \cite{moon2013}. As a result, the current required to make some of the spin-wave energies negative is smaller . 

In conclusion, we have presented a simple set-up and model for spin-wave amplification and lasing by inhomogeneous spin-transfer torques. We note that the amplification and lasing that we have discussed occurs already within the linear spin-wave approximation. As we have mentioned, the energy of the emitted spin waves is provided by creating negative-energy excitations in the narrow region of the wire. Future research could focus on more accurate modelling of the set-up in Fig.~\ref{fig:setup}, including non-linearities, dipolar interactions and more complicated current patterns. These issues could be addressed by numerical solution of the stochastic Landau-Lifshitz-Gilbert equation. Recently, numerical results were reported that show spin-wave emission in a similar set-up as the one we consider \cite{slavin2018}. A direct comparison between these results and ours is postponed to future work.

{\it Acknowledgements.---} We thank Andrei Slavin for sharing unpublished results and Reinoud Lavrijsen for useful discussions and comments on our manuscript. R.D. is member of the
D-ITP consortium, a program of the Netherlands Organisation
for Scientific Research (NWO) that is funded by
the Dutch Ministry of Education, Culture and Science
(OCW). This work is in part funded by the Stichting voor Fundamenteel
Onderzoek der Materie (FOM) and the European Research Council (ERC).

\end{document}